%% file: letter.tex
\documentclass[aps,prl,reprint,floatfix,superscriptaddress,nofootinbib]{revtex4-2}
\usepackage{amsmath,amssymb,graphicx,booktabs,multirow,hyperref}
\hypersetup{hidelinks}
\newcommand{\FS}{\mathrm{FS}}
\newcommand{\pt}{\mathrm{pt}}
\newcommand{\fold}{\mathrm{fold}}
\newcommand{\TB}{\mathrm{3B}}
\begin{document}
\title{Finite-Size Heavy-Diquarks and State-Dependent Corrections to Heavy Antiquark-Diquark Symmetry}
\author{Guo-Fa Yang}
\affiliation{School of Science, Shenzhen Campus of Sun Yat-sen University, Shenzhen 518107, China}

\author{Tian-Wei Wu}
\email[Corresponding author:] {wutw6@mail.sysu.edu.cn}
\affiliation{School of Science, Shenzhen Campus of Sun Yat-sen University, Shenzhen 518107, China}

\begin{abstract}
Heavy antiquark–diquark symmetry (HADS) maps a compact color-antitriplet heavy pair onto a pointlike antiquark source, but retaining a finite diquark mass does not retain its spatial extent. We develop a finite-size framework that treats the heavy diquark as a composite cluster with an independently determined mass and scale. A unified peak-height-weighted radius prescription applies to ground and radially excited diquarks in nonstrange and strange environments. For 16 selected states of $\Xi_{cc}$, $\Omega_{cc}$, $\Xi_{bb}$, and $\Omega_{bb}$, the mean absolute mass deviation from the central AL1 three-body spectrum falls from 108 MeV in the point-source limit to 10 MeV, without fitting to baryon levels. The structure-resolution angle defines $\delta_{\rm HADS}=1-\langle\cos\hat\theta\rangle$, a quantitative, state-dependent measure of the finite-size contribution to HADS breaking. Internal excitation increases this correction, whereas external excitation reduces it, demonstrating that the same heavy source can be resolved differently by distinct surrounding states. Comparisons across reference potentials preserve these trends and the spectral improvement.

\end{abstract}         
\maketitle

The discovery of tetraquark and pentaquark candidates has extended spectroscopy beyond conventional $q\bar q$ mesons and $qqq$ baryons~\cite{tcc2022,penta2015,BESIII:2013ris,LHCb:2014zfx,LHCb:2020bwg,LHCb:2019kea,LHCb:2022ogu,Liu:2019zoy,Chen:2022asf,Guo:2017jvc,Johnson:2024omq}. The growing configuration space and proliferation of color, spin, and rearrangement channels challenge both resolved calculations and their interpretation~\cite{hiyama2003,richard2018}. As in nuclear cluster descriptions~\cite{Freer:2017gip,Hiyama:2012sma}, correlated quarks may supply effective degrees of freedom~\cite{maiani2005,ebert2002,eakins2012}. Their usefulness requires cluster properties that remain consistent across states and environments.

Heavy-quark symmetries give this cluster picture a physical foundation by identifying properties of a heavy source that its light surroundings cannot resolve. A compact color-antitriplet $QQ$ pair presents the leading color charge of a heavy antiquark, giving heavy antiquark--diquark symmetry (HADS)~\cite{savage1990,brambilla2005,fleming2006,hu2006,mehen2006}, a realization of superflavor symmetry~\cite{Ma:2015lba,Ma:2015cfa,Ma:2017nik}. Replacing one source by the other then relates the spectra and transitions of hadrons with the corresponding light degrees of freedom; this is the basis of applications to doubly heavy baryons, tetraquarks, and heavy-hadron molecules~\cite{eichten2017,mehen2017,guo2013,eakinsdecays2012,shi2020,Wu:2022gie,Meng:2018zbl,Pan:2020xek,Zhang:2025vew}. The observations of $\Xi_{cc}^{++}$, $\Xi_{cc}^{+}$, $\Omega_{cc}^{+}$ and $T_{cc}^{+}$ make systems containing a heavy pair experimentally accessible~\cite{lhcb2017,LHCb:2026lnj,LHCb:2026pxn,LHCb:2021auc,LHCb:2018zpl,LHCb:2019epo,tcc2022}. The fully charmed structures observed in the $J/\psi$-pair spectrum, notably $X(6600)$, $X(6900)$ and $X(7100)$~\cite{LHCb:2020bwg,CMS:2023owd,ATLAS:2023bft,CMS:2025fpt}, further motivate testing heavy-quark clusters in multiquark systems. For the HADS source replacement, however, the spatial structure of the $QQ$ pair must remain unresolved by its surroundings.

For real charm and bottom systems, finite-mass dynamics and finite spatial extent provide distinct, dynamically related departures from this limit~\cite{cohen2006,an2019,mehen2019,soto2020}. Finite masses and calibrated pair interactions enter a resolved constituent calculation. Replacing the heavy pair by a finite-mass constituent retains its internal energy and recoil but erases the two heavy--light distances, allowing a point-cluster reduction to overbind~\cite{richard2018}. This spatial loss matters especially for charm and internal excitations~\cite{cohen2006,eakins2012}.

Finite-size form factors, confinement-geometry corrections, and quark--diquark potentials address aspects of this problem~\cite{ebert2002,Ebert:2007rn,kiselev2018,jido2016,watanabe2022}. A transferable cluster nevertheless requires a common prescription for ground and excited internal states, with unchanged inputs when an external state or light flavor changes. The required scale must therefore characterize the isolated heavy pair, while the finite-size correction to HADS depends on how the external state resolves that pair.

We address this issue through two independent two-body problems solved sequentially: the isolated $QQ$ state supplies a mass and an effective size, which are then used in the external diquark--quark problem. A general finite-size interaction and a structure-resolution angle connect this composite cluster to the point-source limit. We test one scale prescription against ground and radially excited $QQq$ states for $Q=c,b$ and $q=n,s$, where $n$ denotes an isospin-symmetric $u/d$ quark. The same framework then defines an explicitly calculable, state-dependent measure of the finite-size contribution to HADS breaking.

\begin{figure}[t]
\centering
\includegraphics[width=.80\columnwidth]{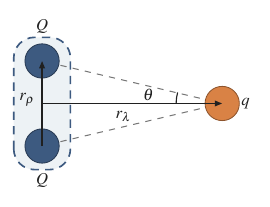}
\caption{Finite-size framework: $r_\rho$ is the internal displacement, $r_\lambda$ the external displacement, and $\theta$ the structure-resolution angle between a heavy--light displacement and $\mathbf r_\lambda$. }
\label{fig:geometry}
\end{figure}

For identical heavy quarks, define the internal displacement $\mathbf r_\rho=\mathbf r_{Q_1}-\mathbf r_{Q_2}$ and the external displacement $\mathbf r_\lambda=\mathbf r_q-(\mathbf r_{Q_1}+\mathbf r_{Q_2})/2$. Let $v(r)$ denote the light--source interaction when the heavy quarks coincide, including the sum of the two pair terms and their color factors. The resolved interaction is
\begin{equation}
 \begin{aligned}
 U(\mathbf r_\lambda,\mathbf r_\rho)&=\tfrac12[v(r_+)+v(r_-)],\\
 r_\pm&=|\mathbf r_\lambda\pm\mathbf r_\rho/2|.
 \end{aligned}
 \label{eq:geometry}
\end{equation}
The isolated internal state $\phi_{n_\rho}$ has energy $\epsilon_{n_\rho}$ and mass $m_{D,n_\rho}=2m_Q+\epsilon_{n_\rho}$. Its mass is used in the external problem, with reduced mass $\mu_{n_\rho}=m_{D,n_\rho}m_q/(m_{D,n_\rho}+m_q)$. For a full heavy-pair separation $r_{QQ,n_\rho}$, the transverse representative geometry of Fig.~\ref{fig:geometry} gives
\begin{equation}
 \begin{aligned}
 s_{n_\rho}(r_\lambda)&=\sqrt{r_\lambda^2+r_{QQ,n_\rho}^2/4},\\
 V_{\FS,n_\rho}(r_\lambda)&=v\!\left[s_{n_\rho}(r_\lambda)\right].
 \end{aligned}
 \label{eq:fs}
\end{equation}
The point source uses $V_{\pt}(r_\lambda)=v(r_\lambda)$. Both prescriptions use the composite kinetic mass $m_{D,n_\rho}$ and give $M=m_{D,n_\rho}+m_q+E_{n_\rho n_\lambda}$; their difference isolates spatial extent at fixed mass input. Equation~\eqref{eq:fs} is a representative-configuration reduction, not an exact angular average. Its geometry is model independent for local central pair interactions, while the mass, scale, and wave function are dynamical inputs.

We determine the scale from the normalized isolated radial probability $\mathcal P_{n_\rho}(r_\rho)=|u_{n_\rho}(r_\rho)|^2$, with $\int_0^\infty\mathcal P_{n_\rho}(r_\rho)\,dr_\rho=1$. If $r_i$ are its local maxima, define
\begin{equation}
 r_{QQ,n_\rho}=\frac{\sum_i\mathcal P_{n_\rho}(r_i)r_i}
 {\sum_i\mathcal P_{n_\rho}(r_i)}.
 \label{eq:scale}
\end{equation}
For the ground state this reduces to the most-probable separation. For excited states it retains every radial lobe through its peak height. This is one prescription for all $n_\rho$: the radius changes with internal excitation, but neither $m_{D,n_\rho}$ nor $r_{QQ,n_\rho}$ is readjusted when the external state or light flavor changes. The isolated distribution fixes both inputs before any three-body mass is consulted.

AL1 is the primary numerical model: we use its spin-independent interaction $v(r)=-A/r+\sigma r-D$ with established masses and color factors~\cite{richard2018}. Here $A=100.025$ MeV fm, $\sigma=837.696$ MeV/fm, $D=832.1$ MeV, and $(m_n,m_s,m_c,m_b)=(315,577,1836,5227)$ MeV. The antitriplet $QQ$ potential is $v/2$. Gaussian expansion~\cite{hiyama2003} solves the two-body problems and the three-body benchmark with all three Jacobi rearrangements in the implemented $S$-wave basis. Hyperfine, spin--orbit, and tensor interactions are omitted.

We compare $(1,1)$, the external excitation $(1,2)$, and internal excitations $(2,1)$ and $(3,1)$. Their three-body counterparts are identified by the radial nodes along both coordinates, rather than by mass matching alone, with wave-function projections providing a consistency check. Table~\ref{tab:masses} gives the mass order after this structural assignment; these labels identify whether radial excitation occurs predominantly within the heavy pair or between the pair and the light quark, rather than exact quantum numbers~\cite{padmanath2015,eakins2012}.

\begin{figure}[t]
\centering
\includegraphics[width=\columnwidth]{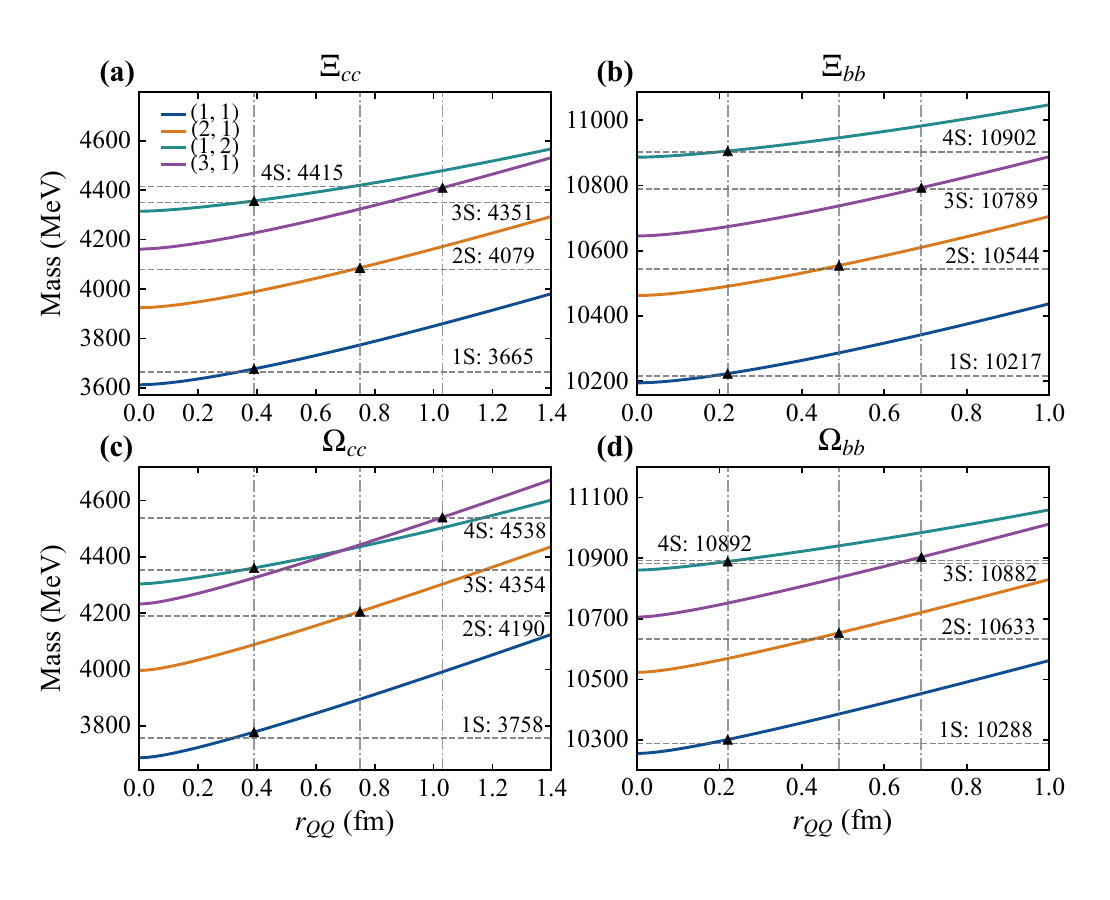}
\caption{AL1 finite-size baryon masses versus the full heavy-pair scale $r_{QQ}$. Panels (a)--(d) are $\Xi_{cc}$, $\Xi_{bb}$, $\Omega_{cc}$, and $\Omega_{bb}$. Solid curves show the radial states $(n_\rho,n_\lambda)$; horizontal dashed lines give the matched three-body masses. Vertical dash-dotted lines mark $r_{QQ,n_\rho}$ from Eq.~\eqref{eq:scale}; triangles give the corresponding finite-size masses.}
\label{fig:scan}
\end{figure}

Figure~\ref{fig:scan} directly tests the scale. At fixed internal mass, the finite-size masses increase monotonically with $r_{QQ}$ because $v'(r)>0$. Each matched three-body level has one crossing on the displayed curve, and the independently determined peak-weighted scale lies nearby for all 16 states. The same internal scale accommodates both external excitations and both light flavors.

For the first three AL1 $cc$ states, the peak-weighted scales are 0.39, 0.75, and 1.03 fm, below the mean separations 0.49, 0.95, and 1.31 fm and rms separations 0.54, 1.03, and 1.43 fm. The latter choices raise the spectrum away from the crossings. Peak heights measure local concentration, whereas lobe probabilities also reflect their widths. In the second $cc$ state the two normalized peak heights are $(0.361,0.639)$, while the lobe probabilities are $(0.206,0.794)$. The broader outer lobe receives less weight in Eq.~\eqref{eq:scale} than in a probability-area average and explains which spatial configurations the scale favors.

\begin{figure*}[t]
\centering
\includegraphics[width=.90\textwidth]{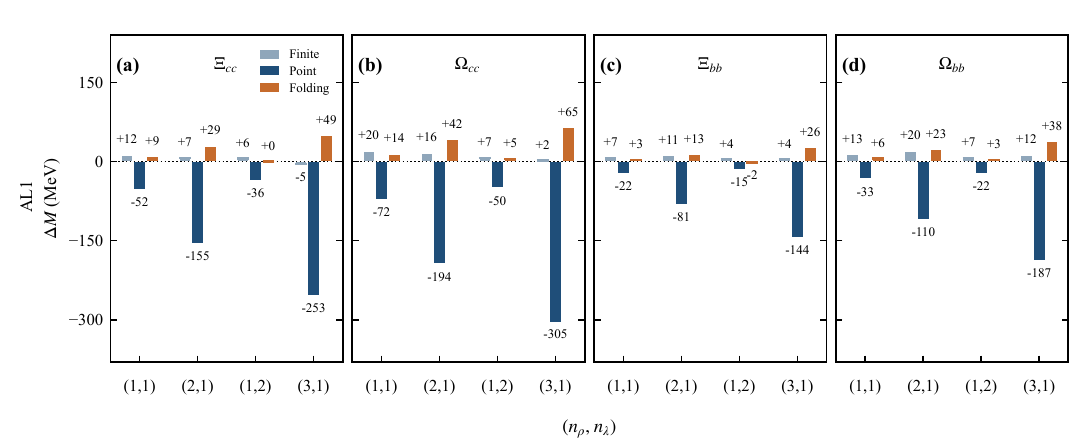}
\caption{AL1 residuals $\Delta M=M_{Dq}-M_{\TB}$ for finite-size, point-source, and folded potentials. Columns are $\Xi_{cc}$, $\Omega_{cc}$, $\Xi_{bb}$, and $\Omega_{bb}$. The dotted line marks zero.}
\label{fig:residuals}
\end{figure*}

For comparison, we also average the pair interaction over the frozen isolated density, a static implementation of the finite-size information encoded in diquark form-factor approaches~\cite{ebert2002}:
\begin{equation}
 V_{\fold,n_\rho}(r_\lambda)=
 \langle\phi_{n_\rho}|U(\mathbf r_\lambda,\mathbf r_\rho)|\phi_{n_\rho}\rangle_\rho.
 \label{eq:fold}
\end{equation}
The folded potential retains the isolated heavy pair's full radial distribution and averages its orientations. Like Eq.~\eqref{eq:fs}, it keeps that internal distribution fixed when the light quark is added. Table~\ref{tab:masses} and Fig.~\ref{fig:residuals} compare the three reductions.

\begin{table}[h]
\caption{AL1 central-potential masses (MeV). The 1S--4S labels give three-body mass order; $(n_\rho,n_\lambda)$ assignments use radial nodes, with wave-function projections as a consistency check.}
\label{tab:masses}
\centering\scriptsize\setlength{\tabcolsep}{2pt}\renewcommand{\arraystretch}{0.92}
\input{tables/table2_masses.tex}
\end{table}

All 16 point-source masses lie below their three-body matches, with a deficit reaching 305 MeV in $\Omega_{cc}(3,1)$ despite including the internal excitation in $m_D$. Finite size reduces the mean absolute residual from 108.2 to 9.6 MeV, with individual residuals between $-5$ and $+20$ MeV. Since $s\geq r_\lambda$ and $v'>0$, $V_{\FS}\geq V_{\pt}$ explains the upward shift. In $\Xi_{cc}$ it is 64, 42, and 248 MeV for $(1,1)$, $(1,2)$, and $(3,1)$, respectively: internal energy alone cannot generate this pattern.

Folding gives a mean absolute residual of 20.4 MeV and is not uniformly superior. Its $\Xi_{cc}(3,1)$ and $\Omega_{cc}(3,1)$ residuals are $+49$ and $+65$ MeV, against $-5$ and $+2$ MeV for finite size. Keeping more information about the isolated density therefore does not by itself ensure a better effective interaction for the baryon. This contrast can be understood by separating radial and angular reduction. Write $a=n_\rho$, $n=n_\lambda$, $R=r_\lambda$, and $r=r_\rho$, and introduce the radially averaged transverse potential
\begin{equation}
V_{\perp,a}(R)=\int_0^\infty dr\,\mathcal P_a(r)
 v\!\left(\sqrt{R^2+r^2/4}\right).
\label{eq:transverse}
\end{equation}
Denote the expectation values of these two successive changes in the same finite-size external state by $I_{\rm rad}$ and $I_{\rm ang}$, respectively. Their sum $B=I_{\rm rad}+I_{\rm ang}=\langle V_{\fold}-V_{\FS}\rangle$ is the frozen-interaction correction:
\begin{equation}
\begin{aligned}
 I_{\rm rad}&=\langle V_{\perp,a}-V_{\FS,a}\rangle_n,\\
 I_{\rm ang}&=\langle V_{\fold,a}-V_{\perp,a}\rangle_n,\\
 B_{an}&\equiv\langle V_{\fold,a}-V_{\FS,a}\rangle_n
             =I_{\rm rad}+I_{\rm ang}.
\end{aligned}
\label{eq:interaction_reduction}
\end{equation}
The peak scale must be understood together with the transverse geometry. Two changes distinguish this prescription from folding: replacing the radial distribution by one representative separation, and replacing the orientation average by a transverse configuration. 
For the tested states, restoring the radial distribution at transverse geometry raises the interaction relative to the peak-scale value. 
Conversely, averaging orientations lowers the transverse interaction: for the Coulomb-plus-linear potential, $v(\sqrt t)$ is concave, so this direction follows from Jensen's inequality~\cite{jensen1906}. In $\Xi_{cc}(1,1)$, $I_{\rm rad}=+29.8$ MeV and $I_{\rm ang}=-33.7$ MeV give $B=-3.9$ MeV. These two changes partly cancel, particularly for ground and externally excited states. The peak-weighted radius thus selects the most concentrated radial configurations without importing the full weight of a broad outer tail, while the representative geometry supplies a compensating angular contribution.

An additional effect appears when the heavy pair is embedded in the baryon. The two heavy--light interactions couple $r_\rho$ and $r_\lambda$, so the internal distribution and the light-quark wave function can readjust together instead of remaining a product of a fixed isolated internal state and an external state. 
Introduce the radial dynamical response 
\begin{equation}
\Delta E_{{\rm rad},an}=E^{\rm rad}_{k(a,n)}-\langle \Phi_{an}|H_3|\Phi_{an}\rangle,
\label{eq:radial_response}
\end{equation}
with $\Phi_{an}=\phi_a(r_\rho)\chi^{\rm FS}_{an}(r_\lambda)$, where $\phi_a$ is an isolated diquark eigenstate and
$\chi^{\rm FS}_{an}$ is the corresponding external eigenstate
of the finite-size potential. The energy $E^{\rm rad}_{k(a,n)}$
is obtained by diagonalizing the original three-body Hamiltonian
$H_3$ in the space spanned by internal--external radial product
states with $\ell_\rho=\ell_\lambda=0$, selecting the eigenstate
corresponding to the target excitation.
Thus, $\Delta E_{{\rm rad},an}$ includes both internal radial
deformation and external-wave-function readjustment relative
to the frozen product state. 
For $\Xi_{cc}(3,1)$, $I_{\rm rad}=+106.3$ MeV and $I_{\rm ang}=-52.4$ MeV leave $B=+54.0$ MeV, largely offset by $\Delta E_{\rm rad}=-48.2$ MeV. The corresponding $\Omega_{cc}(3,1)$ contributions are $B=+62.3$ MeV and $\Delta E_{\rm rad}=-62.1$ MeV. This diagnostic uses $2m_Q$ in the external Jacobi kinetic energy instead of the cluster mass $m_D$; the associated kinetic expectation-value changes are $+2.0$ and $+2.7$ MeV, respectively. In these internally excited charm states, radial readjustment therefore offsets much of the upward shift from restoring the frozen pair interaction. The finite-size potential does not calculate this response explicitly; its smaller representative radius can effectively compensate part of the missing energy change. Thus there are two distinguishable contributions to
the success of the effective prescription: radial–angular compensation in the interaction, and dynamical internal
reorganization. This explains why it can reproduce masses more closely than frozen-density folding. The evidence supports the combined scale-and-geometry prescription as an effective spectral description.

\begin{table}[t]
\caption{AL1 structure properties. $\theta=\langle\hat\theta\rangle$ is in degrees, $\delta_{\rm HADS}=\langle\hat\delta\rangle$ in percent, and rms radii in fm. \textit{Finite} and \textit{3B} denote finite-size and full three-body results.}
\label{tab:structure}
\centering\footnotesize\setlength{\tabcolsep}{2pt}\renewcommand{\arraystretch}{0.96}
\input{tables/table3_structure.tex}
\end{table}

The same geometry supplies a quantitative measure of the state-dependent HADS correction. The angle between either heavy--light displacement and $\mathbf r_\lambda$ in the representative configuration defines a coordinate-space multiplication operator
\begin{equation}
 \begin{aligned}
 \hat\theta(r_\lambda)&=\arctan\frac{r_{QQ}}{2r_\lambda},\\
\cos\hat\theta(r_\lambda)&=\frac{r_\lambda}{s_{n_\rho}(r_\lambda)}.
 \end{aligned}
 \label{eq:angle}
\end{equation}
For the normalized external radial state $u^{\FS}_{n_\rho n_\lambda}$, the reported angle is $\theta=\langle\hat\theta\rangle_{n_\rho n_\lambda}$. Define the finite-size characteristic operator $\hat\delta=1-\cos\hat\theta$ and its expectation value
\begin{equation}
 \begin{aligned}
 \delta_{{\rm HADS}}
 &=1-\langle\cos\hat\theta\rangle_{n_\rho n_\lambda}\\
 &=\int_0^\infty dr_\lambda\,|u^{\FS}_{n_\rho n_\lambda}(r_\lambda)|^2\times\left[1-\frac{r_\lambda}{s_{n_\rho}(r_\lambda)}\right].
 \end{aligned}
 \label{eq:delta}
\end{equation}
Bounded between zero and one and vanishing for a point source, this expectation value quantifies how the external state resolves the finite heavy source. Its geometrical definition is independent of the chosen potential; its scale and wave function are dynamical. It provides a quantitative finite-size HADS correction within the framework, rather than a percentage of all symmetry-breaking effects. Its spectral connection follows exactly at fixed $m_D,m_q$ from the Hellmann--Feynman theorem~\cite{feynman1939}:
\begin{equation}
\frac{\partial E^{\FS}}{\partial\ln r_{QQ}}
=\left\langle s v'(s)(1+c)(1-c)\right\rangle,\quad c=\frac{r_\lambda}{s}.
\label{eq:HF}
\end{equation}
The factor $1-c$ is the local finite-size resolution measured by $\delta_{\rm HADS}$ in Eq.~\eqref{eq:delta}; Eq.~\eqref{eq:HF} weights that same factor by $s v'(s)(1+c)$ to obtain the energy sensitivity $\partial E^{\FS}/\partial\ln r_{QQ}$. Thus the structural correction and the spectral sensitivity have an explicit dynamical connection, although they are different quantities with different units. Neither $\delta_{\rm HADS}$ nor this energy derivative is an estimate of the residual error relative to the three-body spectrum.

\begin{figure}[h]
\centering
\includegraphics[width=.45\textwidth]{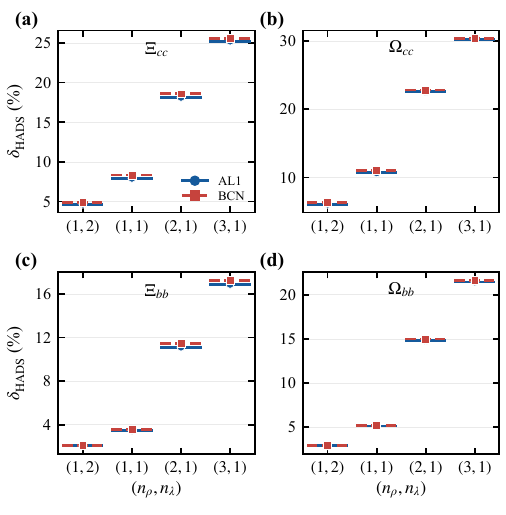}
\caption{State-dependent $\delta_{\rm HADS}$ (\%) for the primary AL1 model (blue circles), with BCN (red squares) as a reference comparison. Columns show $\Xi_{cc}$, $\Omega_{cc}$, $\Xi_{bb}$, and $\Omega_{bb}$; states are ordered as $(1,2)$, $(1,1)$, $(2,1)$, and $(3,1)$.}
\label{fig:state}
\end{figure}

State comparisons separate source input from spatial resolution (Table~\ref{tab:structure} and Fig.~\ref{fig:state}). From $(1,1)$ to $(1,2)$, $m_D$, $r_{QQ}$, and the reduced mass are unchanged. The $\Xi_{cc}$ external rms radius grows from 0.69 to 1.25 fm while $\delta_{\rm HADS}$ falls from 7.9\% to 4.7\%; in $\Xi_{bb}$ it falls from 3.5\% to 2.1\%. The same finite-mass cluster appears more pointlike to an extended external state.

Internal excitation reverses the trend. Along $\Xi_{cc}(1,1),(2,1),(3,1)$, $\delta_{\rm HADS}$ rises from 7.9\% to 18.1\% and 25.2\%, while the external radius changes little. The corresponding bottom values, 3.5\%, 11.1\%, and 16.9\%, reflect a more compact source. Each point-to-finite-size comparison still fixes its own internal mass. Heavy-quark mass alone therefore does not specify this correction.

Changing $n$ to $s$ holds the heavy-cluster inputs fixed but increases the external reduced mass and favors shorter distances. Ground-state $\delta_{\rm HADS}$ then grows from 7.9\% to 10.7\% for $cc$ and from 3.5\% to 5.1\% for $bb$. Thus internal structure and external environment jointly control the correction. All four systems share $(1,2)<(1,1)<(2,1)<(3,1)$ for both $\theta$ and $\delta_{\rm HADS}$.

Reference interactions BCN, AL2, AP1, AP2, and BD~\cite{genovese1999,Silvestre-Brac:1996myf} preserve the spectral improvement and these structural orderings when the same isolated-scale prescription is applied. For the 16 states considered, the average finite size residual is 12.4 MeV, which remains superior to the point (112.6 MeV) and folding (23.4 MeV) cases, the state-by-state spread in $\delta_{\rm HADS}$ across these potentials ranges from 0.3 to 1.3 percentage points. They test robustness within related central constituent models. This supports
the finite-size prescription as a physically specified cluster
input with consistency across states and environments.

The finite-size calculation fixes one isolated internal state $\phi_{n_\rho}$ and its radius while solving the external motion. At higher excitation, internal and external excitations can mix appreciably in a three-body eigenstate. A single $(n_\rho,n_\lambda)$ assignment and radius may then be insufficient, requiring a coupled three-body treatment. This limits the fixed-internal-state approximation: $\delta_{\rm HADS}$ remains well defined for each chosen cluster state, but one such value cannot characterize a strongly mixed three-body state.

A heavy diquark with an independently determined mass and peak-weighted scale provides an effective constituent for the selected doubly heavy baryons. The same isolated-cluster inputs remove most of the point-source overbinding without readjustment when the external state or light flavor changes. The structure-angle operator connects this finite-size constituent to the external state's spatial resolution, giving a quantitative, state-dependent finite-size correction to HADS. The correction thus depends on both the heavy source and its surroundings, rather than on heavy-quark mass alone. Applying the same cluster inputs to doubly heavy and fully heavy tetraquarks would test their transferability when the surrounding constituents and color couplings change. Spin splittings and transition observables provide further tests of the resolution measure beyond the mass spectrum. These extensions are needed to establish the predictive range of the finite-size cluster description.

This work is supported by the National Natural Science Foundation of China (NSFC) under Grant No.12405108. Tian-Wei Wu acknowledges the support from the Appointed Position of Pengcheng Peacock Plan (C Track), Shenzhen.

\bibliographystyle{apsrev}
\bibliography{references}
\end{document}

%% file: tables/table2_masses.tex
\begin{tabular}{ccccccc}
\hline\hline
System & $(n_\rho,n_\lambda)$ & State & \shortstack{Point} & \shortstack{Finite} & \shortstack{Fold} & \shortstack{3-body}\\
\midrule
\multirow{4}{*}{$\Xi_{cc}$} & $(1,1)$ & 1S & 3613 & 3677 & 3674 & 3665\\
 & $(2,1)$ & 2S & 3924 & 4086 & 4108 & 4079\\
 & $(1,2)$ & 3S & 4315 & 4357 & 4351 & 4351\\
 & $(3,1)$ & 4S & 4162 & 4410 & 4464 & 4415\\
\midrule
\multirow{4}{*}{$\Omega_{cc}$} & $(1,1)$ & 1S & 3686 & 3778 & 3772 & 3758\\
 & $(2,1)$ & 2S & 3996 & 4206 & 4232 & 4190\\
 & $(1,2)$ & 3S & 4304 & 4361 & 4359 & 4354\\
 & $(3,1)$ & 4S & 4233 & 4540 & 4603 & 4538\\
\midrule
\multirow{4}{*}{$\Xi_{bb}$} & $(1,1)$ & 1S & 10195 & 10224 & 10220 & 10217\\
 & $(2,1)$ & 2S & 10463 & 10555 & 10557 & 10544\\
 & $(3,1)$ & 3S & 10645 & 10793 & 10815 & 10789\\
 & $(1,2)$ & 4S & 10887 & 10906 & 10900 & 10902\\
\midrule
\multirow{4}{*}{$\Omega_{bb}$} & $(1,1)$ & 1S & 10255 & 10301 & 10294 & 10288\\
 & $(2,1)$ & 2S & 10523 & 10653 & 10656 & 10633\\
 & $(1,2)$ & 3S & 10860 & 10889 & 10885 & 10882\\
 & $(3,1)$ & 4S & 10705 & 10904 & 10930 & 10892\\
\hline\hline
\end{tabular}

%% file: tables/table3_structure.tex
\begin{tabular}{cccccc}
\hline\hline
System & $(n_\rho,n_\lambda)$ & $\theta$ & \shortstack{$\delta_{\rm HADS}$} & $r_{\lambda,\rm rms}^{\rm Finite}$ & \shortstack{$(r_{\rho,\rm rms}^{3B},$$r_{\lambda,\rm rms}^{3B})$}\\
\midrule
\multirow{4}{*}{$\Xi_{cc}$} & $(1,1)$ & 21 & 7.9 & 0.69 & (0.48, 0.68)\\
 & $(2,1)$ & 33 & 18.1 & 0.73 & (0.89, 0.73)\\
 & $(1,2)$ & 14 & 4.7 & 1.25 & (0.51, 1.24)\\
 & $(3,1)$ & 40 & 25.2 & 0.75 & (1.19, 0.77)\\
\midrule
\multirow{4}{*}{$\Xi_{bb}$} & $(1,1)$ & 13 & 3.5 & 0.66 & (0.30, 0.66)\\
 & $(2,1)$ & 25 & 11.1 & 0.69 & (0.60, 0.68)\\
 & $(1,2)$ & 9 & 2.1 & 1.22 & (0.31, 1.21)\\
 & $(3,1)$ & 32 & 16.9 & 0.71 & (0.83, 0.71)\\
\midrule
\multirow{4}{*}{$\Omega_{cc}$} & $(1,1)$ & 24 & 10.7 & 0.57 & (0.47, 0.55)\\
 & $(2,1)$ & 38 & 22.5 & 0.61 & (0.86, 0.61)\\
 & $(1,2)$ & 16 & 6.1 & 1.04 & (0.50, 1.02)\\
 & $(3,1)$ & 44 & 30.2 & 0.64 & (1.17, 0.66)\\
\midrule
\multirow{4}{*}{$\Omega_{bb}$} & $(1,1)$ & 16 & 5.1 & 0.53 & (0.29, 0.52)\\
 & $(2,1)$ & 30 & 14.8 & 0.56 & (0.58, 0.55)\\
 & $(1,2)$ & 10 & 2.9 & 0.99 & (0.41, 0.94)\\
 & $(3,1)$ & 36 & 21.5 & 0.58 & (0.76, 0.65)\\
\hline\hline
\end{tabular}